\documentstyle [12pt,aasms4]{article}

\def\la{\mathrel{\hbox{\rlap{\hbox{\lower4pt\hbox{$\sim$}}}\hbox{$<$}}}}
\def\ga{\mathrel{\hbox{\rlap{\hbox{\lower4pt\hbox{$\sim$}}}\hbox{$>$}}}}

\begin{document}
\title{The physics of crystallizing white dwarfs}

\author{J. Isern\altaffilmark{1},
	R. Mochkovitch\altaffilmark{2},
	E. Garc\'{\i}a-Berro\altaffilmark{3},
	\and
	M. Hernanz\altaffilmark{1}} 

\altaffiltext{1}{Institute for Space Studies of Catalonia (CSIC 
		 Research Unit), Edifici Nexus--104, Gran Capit\`a 
		 2--4, 08034 Barcelona, Spain}
\altaffiltext{2}{Institut d'Astrophysique de Paris, CNRS, 98bis Bd 
		 Arago, 75014 Paris, France}
\altaffiltext{3}{Departament de F\'{\i}sica Aplicada, Universitat 
		 Polit\`ecnica de Catalunya, Jordi Girona Salgado 
		 s/n, M\`odul B--5, Campus Nord, 08034 Barcelona, 
		 Spain}

\received{}
\accepted{}

\begin{abstract}
White dwarfs can be used as galactic chronometers and, therefore, 
provide important information about galactic evolution if good
theoretical models of their cooling are available. Consequently, 
it is natural to wonder if all the sources or sinks of energy 
are correctly taken into account. One of these sources is partial 
differentiation of the chemical components of the white dwarf upon 
crystallization. In this paper we use a new formalism to show that 
if there is a redistribution of the elements inside the star, there 
is a net release of energy that has to be radiated away and that 
slows down the cooling rate of the white dwarf.  
\end{abstract}

\keywords{stars: interiors --- white dwarfs.}

\section {Introduction}

The evolution of white dwarfs is essentially a cooling process that 
lasts for $\sim 10$ Gyr. Since the study of white dwarfs allows to 
obtain information about the age of the Galaxy (Winget et al. 1987, 
Garc\'{\i}a-Berro et al. 1988a,b, Hernanz et al. 1994), it is important 
to identify all the sources of energy as well as the mechanisms that 
control its outflow.

The vast majority of white dwarfs (those with masses in the range 
$0.45 \la M/M_{\sun} \la 1.1$) are made of a mixture of carbon, 
oxygen and some impurities coming from the metal content 
of the parent star. The most important of these impurities is 
$^{22}$Ne, which results from He-burning on the ashes of the 
CNO cycle, and reaches $\sim$ 2\% by mass in Population I 
stars. Since during the cooling process the star experiences a 
phase transition, it is natural to wonder if a change of solubility 
at the onset of crystallization can provide an extra source of energy 
(Schatzman 1958, Stevenson 1980, Mochkovitch 1983).

Segretain and Chabrier (1993) and Segretain et al. (1994) computed 
phase diagrams for arbitrary binary mixtures in terms of the 
modern density-functional theory of freezing. They showed that 
the shape of the phase diagram was completely characterized by 
the charge ratio of the mixture, $Z_1/Z_2$. Their diagrams evolve 
from the spindle form for $0.72 \leq Z_1/Z_2 < 1$, into an azeotropic 
form for $0.58 \leq Z_1/Z_2 < 0.72$ and finally into an eutectic form
for $Z_1/Z_2 < 0.58$. Using these phase diagrams, Hernanz et al. 
(1994) showed that the chemical differentiation of pure carbon-oxygen 
mixtures could introduce a delay of up to $\sim 2$ Gyr, depending on 
the adopted initial profiles of carbon and oxygen in the white dwarf 
before freezing. The settling of $^{22}$Ne was much more spectacular 
since it could keep the white dwarf warm during several billion years, 
if the abundance of neon was assumed to be $X_{22}=0.02$ (Isern et 
al. 1991, Segretain et al. 1994). The deposition of $^{56}$Fe, the 
second impurity in importance, turned out to be less important (Xu 
and Van Horn 1992, Segretain et al. 1994) because of its smaller 
abundance, $X_{56}=0.001$. In any case, and contrary to the 
carbon-oxygen case, these impurities should not affect the 
determination of the disk age from white dwarfs since their 
abundances are expected to be negligible in the oldest objects 
(Hernanz et al. 1994).

Although the importance of the mechanical and thermodynamical 
consequences of the solidification of alloys has been recognized 
in Geophysics (Loper 1978, Chen 1995), this is not the case in 
Astrophysics and very often these phenomena are either completely 
ignored or sometimes, misinterpreted. In this paper we reexamine 
the efficiency of convective mixing, which is responsible for the 
redistribution of the chemical elements during 
crystallization, and we provide a suitable formalism to compute 
the subsequent energy released during this process.

\section{The physics of the crystallization process}

Due to the spindle shape of the phase diagram of C/O mixtures, 
the solid formed upon crystallization is richer in oxygen than 
the liquid and therefore denser. Using the condition of pressure 
continuity, the density excess can be estimated to be:
\begin{equation}
\frac{\delta \rho}{\rho} \simeq -\frac{\delta P_{\rm i}}{\gamma 
P_{\rm e}} -\frac{\delta Y_{\rm e}}{Y_{\rm e}}
\end{equation}
where $P_{\rm i}$ and $P_{\rm e}$ are the ionic and electronic 
pressures respectively, $\gamma$ is the electron adiabatic index 
and $Y_{\rm e}$ is the number of electrons per nucleon. For a $0.6\, 
M_{\sun}$ white dwarf with equal amounts of carbon and oxygen, $\delta 
\rho /\rho \simeq 10^{-4}$. Therefore, as the solid core grows from 
the center of the star the lighter liquid left behind can be 
redistributed by Rayleigh-Taylor instabilities (Stevenson, 1980; 
Mochkovitch 1983). 

The local energy budget of the white dwarf can be written as:
\begin{equation}
\frac{dL_{\rm r}}{dm}= -\epsilon_\nu -T\frac{dS}{dt}
\end{equation}
where all the symbols have their usual meaning. Assume now that the 
white dwarf is  made of two chemical species with atomic numbers $Z_0$
and $Z_1$, mass numbers $A_0$ and $A_1$, and abundances by mass $X_0$ 
and $X_1$, respectively $(X_0+X_1=1)$. In the following, the suffix 
0 will always refer to the heavier component. Using the First Principle
of Thermodynamics, the right hand side of equation (2) can be written 
as:
\begin{equation}
\frac{dL_{\rm r}}{dm}= -\epsilon_\nu -P\,\frac{dV}{dt}-\frac{dE}{dt}
\end{equation}
where $E$ is the internal energy per unit mass and $V=1/\rho$. We can 
also write
\begin{equation}
\frac{dE}{dt}=
      \Big(\frac{\partial E}{\partial T}\Big)_{V,X_0}\frac{dT}{dt}+
      \Big(\frac{\partial E}{\partial V}\Big)_{T,X_0}\frac{dV}{dt}+
      \Big(\frac{\partial E}{\partial X_0}\Big)_{T,V}\frac{dX_0}{dt}
\end{equation}
Using the elementary thermodynamic relation
\begin{equation}
\Big(\frac{\partial E}{\partial V}\Big)_{T,X_0}=
-P+T\Big(\frac{\partial P}{\partial T}\Big)_{V,X_0}
\end{equation}
equation (2) can now be written as:
\begin{equation}
-(\frac{dL_{\rm r}}{dm}+\epsilon_{\nu})=C_{\rm v}\frac{dT}{dt}+
T\Big(\frac{\partial P}{\partial T}\Big)_{V,X_0}\frac{dV}{dt}
-l_{\rm s}\frac{{dM}_{\rm s}}{dt}\delta(m-M_{\rm s})+
\Big(\frac{\partial E}{\partial X_0}\Big)_{T,V}\frac{X_0}{dt}
\end{equation}
where $l_{\rm s}$ is the latent heat of crystallization and $\dot{M}_
{\rm s}$ is the rate at which the solid core grows; the delta function
indicates that the latent heat is released at the solidification front. 
Notice that chemical differentiation contributes to the luminosity 
not only through compressional work, which is negligible, but also 
through the change in the chemical abundances, which leads to the 
last term of this equation. Notice, as well, that the largest 
contribution to $L_{\rm r}$ due to the change in $E$ exactly
cancels out the $P\,dV$ work for {\sl any} evolutionary change
(with or without a compositional change). This is, of course, 
a well known result (Mestel 1952, Kovetz and Shaviv 1975, Lamb 
and Van Horn 1975, D'Antona and Mazzitelli 1990) that can be 
related to the release of gravitational energy (see below).

\newpage

Finally, integrating (6) over the whole star, we obtain:
\begin{eqnarray}
L+L_{\nu}&=&- \int^{M_{\rm WD}}_0 C_{\rm v}\frac{dT}{dt}\,dm
          - \int^{M_{\rm WD}}_0 T\Big(\frac{\partial P}{\partial 
             T}\Big)_{V,X_0}\frac{dV}{dt}\,dm \nonumber\\
          &+& l_{\rm s} \frac{dM_{\rm s}}{dt}
          - \int^{M_{\rm WD}}_0 \Big(\frac{\partial E}{\partial 
             X_0}\Big)_{T,V}\frac{dX_0}{dt}\,dm
\end{eqnarray}

The first term of equation (7) is the well known contribution of the
heat capacity of the star to the total luminosity (Mestel 1952). The
second term represents the contribution to the luminosity due to the
change of volume. It is in general small since only the thermal part
of the electronic pressure, the ideal part of the ions and the Coulomb
terms other than the Madelung term contribute (Kovetz and Shaviv 1975,
Lamb and Van Horn 1975). However, when the white dwarf enters into the
Debye regime, this term provides about the 80\% of the total luminosity
preventing the sudden disappearence of the star (D'Antona and
Mazzitelli 1990). The third term represents the contribution of
the latent heat to the total luminosity at freezing. Since the
latent heat of Coulomb plasmas is small ($\sim kT_{\rm s}/{\rm
nucleus}$, where $T_{\rm s}$ is the solidification temperature),
its contribution to the total luminosity is modest although not
negligible. The fourth term represents the energy released by the
chemical readjustement of the white dwarf, i.e. the release of the
energy stored in the form of chemical potentials. This term is
usually negligible in normal stars, since it is much smaller than
the energy released by nuclear reactions, but it must be taken into
account when all other energy sources are small.

The last term in equation (7) can be further expanded. Consider 
that the crystallization front is at mass coordinate $M_{\rm s}$,
that in a time interval $\delta t$ the crystallization front 
advances by $\delta M_{\rm s}$, and that the extension of the 
convective shell driven by the Rayleigh-Taylor instability is 
$\Delta M$. Then the change in the chemical abundance of the 
mixing region $(\delta X_0^{\rm liq})$ can be related (assuming 
perfect mixing, see \S 3 below) to the difference between the 
chemical abundances of the liquid and the solid: 
\begin{equation}
\delta M_{\rm s}\,(X_0^{\rm sol}-X_0^{\rm liq})=
-\Delta M\,\delta X_0^{\rm liq} 
\end{equation}
Besides, we can write
\begin{equation}
\int^{M_{\rm WD}}_0 \Big(\frac{\partial E}{\partial X_0}\Big)_{T,V}
\delta X_0\,dm=\Big(\frac{\partial E}{\partial X_0}\Big)_{M_{\rm s}}
\,(X_0^{\rm sol}-X_0^{\rm liq})\,\delta M_{\rm s}+ \delta X_0^{\rm liq}
\,\int_{\Delta M}\Big(\frac{\partial E}{\partial X_0}\Big)_{T,V}
\,dm
\end{equation}
where $(\partial E/\partial X_0)_{M_{\rm s}}$ is the partial derivative 
evaluated at the edge of the solid core. Taking into account equation 
(8), we can now write
\begin{equation}
\int^{M_{\rm WD}}_0 \Big(\frac{\partial E}{\partial X_0}\Big)_{T,V}
\frac{dX_0}{dt}\,dm=
(X_0^{\rm sol}-X_0^{\rm liq})\,\Bigg[\Big(\frac{\partial E}
{\partial X_0}\Big)_{M_{\rm s}}-\Big\langle\frac{\partial E}
{\partial X_0}\Big\rangle\Bigg]\frac{dM_{\rm s}}{dt}
\end{equation}
where
\begin{equation}
\Big\langle\frac{\partial E}{\partial X_0}\Big\rangle=\frac{1}{\Delta 
M}\int_{\Delta M}\Big(\frac{\partial E}{\partial X_0}\Big)_{T,V} dm
\end{equation} 

From equation (10) we can define the total energy released per gram of
crystallized matter due to the change in chemical composition as:
\begin{equation}
\epsilon_{\rm g}=-(X_0^{\rm sol}-X_0^{\rm liq})
\Bigg[\Big(\frac{\partial E}{\partial X_0}\Big)_{M_{\rm s}}-
\Big\langle\frac{\partial E}{\partial X_0}\Big\rangle\Bigg] 
\end{equation}
The square bracket is negative since $(\partial E/\partial X_0)$ is 
negative and essentially depends on the density, which monotonically 
decreases outwards. The internal energy per unit mass can be divided 
into the electronic and the ionic components, so we can write 
$\epsilon_{\rm g}=\epsilon_{\rm e}+\epsilon_{\rm i}$. For the 
sake of simplicity we are only going to use here the completely 
degenerate nonrelativistic expression for the electrons 
\begin{equation}
E_{\rm e}=\frac{3}{2}K_1\rho^{2/3}Y_{\rm e}^{5/3}
\end{equation} 
where $K_1=1.004\times 10^{13}$ (cgs units), and the ideal contribution
plus the Madelung term of the Coulomb energy for the ions
\begin{equation}
E_{\rm i}\simeq \frac{\Re T}{\mu}\Big( \frac{3}{2} -0.9 \Gamma\Big)
\end{equation} 
where $\Re$ is the gas constant and $\Gamma=\Gamma_{\rm e}
\overline{Z^{5/3}}$ is the Coulomb coupling constant, with $\Gamma_
{\rm e}=2.272\times 10^5 (\rho Y_{\rm e})^{1/3}/T$. 

Since the contribution to the electron mole number of carbon and 
oxygen are very similar $\epsilon_{\rm g}$ is dominated by the 
ionic contribution (see \S 4 below). Thus, we can obtain a more 
transparent expression for $\epsilon_{\rm g}$ showing that 
it also corresponds to a release of gravitational energy. We 
first write
\begin{equation}
\epsilon_{\rm g}= -\alpha \Big(X_{\rm 0}^{\rm sol}- X_{\rm 0}^{\rm 
liq}\Big)
\Big(\frac{\partial E}{\partial X_0}\Big)_{M_{\rm s}}
\end{equation}
where we have introduced the parameter
\begin{equation}
\alpha=\frac{\Big(\frac{\partial E}{\partial X_0}\Big)_{M_{\rm s}}- 
\Big\langle\frac{\partial E}{\partial X_0}\Big\rangle}
{\Big(\frac{\partial E}{\partial X_0}\Big)_{M_{\rm s}}}
\la 1
\end{equation}
The ionic pressure associated to the Madelung energy is 
$P_{\rm i}=\frac{1}{3}\rho E_{\rm i}$ so that 
\begin{equation}
\epsilon_{\rm g} \simeq -3 \alpha 
\Big(X_{\rm 0}^{\rm sol}- X_{\rm 0}^{\rm liq}\Big)
\frac{1}{\rho}\Big(\frac{\partial P_{\rm i}}{\partial X_0}\Big)_{M_{\rm 
s}}\simeq -3\alpha\frac{\delta P_{\rm i}}{\rho}= 
3\alpha \gamma\frac{P_{\rm e}}{\rho}\frac{\delta \rho}{\rho}
\end{equation}
$\delta P_{\rm i}$ being the change of ionic pressure at 
crystallization. The last term in the above equation has 
been obtained by taking into account pressure continuity, 
which leads to $\delta P_{\rm i}=-\delta P_{\rm e} \simeq 
-\gamma P_{\rm e} \,{\delta \rho}/{\rho}$. Finally, using 
the virial theorem in the form
\begin{equation}
3\Big\langle \frac{P_{\rm e}}{\rho}\Big\rangle \simeq 
-\frac{\Omega}{M_{\rm WD}} 
\simeq \beta \frac{G M_{\rm WD}}{R_{\rm WD}} \simeq \beta g_{\rm 
WD}R_{\rm WD}
\end{equation}
where the average is taken over the white dwarf mass and where 
$\Omega = - \beta GM_{\rm WD}/R^2_{\rm WD}$, $g_{\rm WD}$ and 
$R_{\rm WD}$ are, respectively, the white dwarf gravitational 
energy, surface gravity and radius ($\beta = 6/7$ for $\gamma 
= 5/3$), we obtain
\begin{equation}
\epsilon_{\rm g} \simeq \alpha\beta\gamma \frac{P_{\rm e}/\rho}
{\big\langle P_{\rm e}/\rho\big\rangle}\,g_{\rm WD}R_{\rm WD}
\frac{\Delta \rho}{\rho} \simeq {\rm k}\,  g_{\rm WD}R_{\rm WD}
\frac{\Delta \rho}{\rho}
\end{equation}
The factor k is of the order of unity except close to the surface of 
the white dwarf where $\rho$ rapidly decreases. With this expression
for $\epsilon_{\rm g}$, it is clearly seen that it corresponds to a 
release of gravitational energy.

\section{The efficiency of the mixing process}

Before computing the energy released by chemical segregation it 
is convenient to check the efficiency of convective mixing. Let 
$v_{\rm crys}$ be the propagation velocity of the solidification 
front into the C/O mixture. This velocity, which can be obtained 
from the models, is very small, $\sim 0.1$ cm/yr. The mass flux 
of carbon released by the front in the liquid phase is:
\begin{equation}
F^{\rm crys}_{\rm C} = \rho v_{\rm crys} (X_{\rm C}^{\rm l}-
X_{\rm C}^{\rm s})=\rho v_{\rm crys}\Delta X_{\rm C}
\end{equation}
where $X_{\rm C}^{\rm l}$ and $X_{\rm C}^{\rm s}$ are the carbon mass 
fractions in the liquid and the solid respectively. The criterion for 
convective instability taking into account heat conduction from the 
convective eddies is (Stevenson and Salpeter 1977):
\begin{equation}
\chi > k\epsilon
\end{equation}
with
\begin{equation}
\chi = -\frac{H_{\rm P}}{\rho c^2_{\rm s}}\,\mu\, \Big(\frac{\partial P}
{\partial\mu}\Big)_{\rho,T}\,\Big(\frac{1}{\mu}\,\frac{d\mu}{dr}\Big)
\end{equation}
\noindent
\begin{equation}
\epsilon = \frac{H_{\rm P}}{\rho c^2_{\rm s}}\,T\,\Big(\frac{\partial P}
{\partial T}\Big)_{\rho,\mu}\,\Bigg[\frac{1}{T}\,\frac{dT}{dr}-
(\Gamma _3-1)\,\frac{1}{\rho}\,\frac{d\rho}{dr}\Bigg]
\end{equation}
\begin{equation}
k = \frac{\tau_{\rm cond}}{\tau_{\rm cond}+\tau_{\rm conv}}
\end{equation}
where $H_{\rm P}$ is the pressure scale height, $c_{\rm s}$ is the 
sound velocity, $\mu$ is the mean molecular weight, $(\Gamma _3 -1)=
(\partial \log T / \partial \log \rho)_{\rm ad}$, and $\tau_{\rm 
cond}$ and $\tau_{\rm conv}$ are the conductive and the convective 
characteristic times, respectively. If $l$ is the mixing length, the 
convective velocity can be written as
\begin{equation}
v_{\rm conv} =c_{\rm s}\, (\chi -k\epsilon)^{1/2}\, \frac{l}{H_{\rm P}}
\end{equation}
and the characteristic times are given by:
\begin{equation}
\tau_{\rm conv}= \frac{l}{v_{\rm conv}}
\end{equation}
\begin{equation}
\tau_{\rm cond}=\frac{l^2}{K_T}
\end{equation}
where $K_T$ is the thermal conductivity. If the carbon mass flux 
released at the crystallization front is to be efficiently carried 
by convection we have
\begin{equation}
F^{\rm cryst}_{\rm C} = F^{\rm conv}_{\rm C} = 
\rho v_{\rm conv} \,\Bigg|\frac{dX_{\rm C}}{dr}\Bigg|\,l
\end{equation}
together with a small superadiabaticity
\begin{equation}
\chi - k\epsilon \sim 0
\end{equation}
If the superadiabaticity is indeed small (which will have to be checked 
on the final results), the gradient of carbon mass fraction in the white
dwarf is given by  
\begin{equation}
\Bigg|\frac{dX_{\rm C}}{dr}\Bigg| =k\,\Bigg|\frac{dX_{\rm C}}{dr}
\Bigg|_{\rm ad}
\end{equation}
where
\begin{equation}
\Bigg|\frac{dX_{\rm C}}{dr}\Bigg|_{\rm ad}=-\frac{48}{\mu^2}\,
\frac{(\Gamma_3-1)\,T\,\Big(\frac{\partial P}{\partial T}\Big)_{\rho 
,\mu}}{\Big(\frac{\partial P}{\partial \mu}\Big)_{\rho,T}}\Bigg|
\frac{1}{\rho}\,\frac{d\rho}{dr}\Bigg|
=
{\rm Q}\Bigg|\frac{1}{\rho} \frac{d\rho}{dr}\Bigg|
\end{equation}
The value of Q has been computed for a 0.6 $M_\odot$ crystallizing C/O 
white dwarf and is typically a few $10^{-2}$. Now it is possible to 
obtain the Peclet number,  ${\rm P}=\tau_{\rm cond}/\tau_{\rm conv}$, 
as 
\begin{equation}
\frac{{\rm P}^2}{{\rm P}+1} \simeq 
\frac{5}{3}\frac{v_{\rm crys}\,\Delta X_{\rm C} H_{\rm P}}{K_T \,{\rm 
Q}} \simeq 0.1\; 
\Bigg(\frac{v_{\rm crys}}{0.1\, {\rm cm/yr}}\Bigg)\; 
\Bigg(\frac{\Delta X_{\rm C}}{0.1}\Bigg)\;
\Bigg(\frac{100 \; {\rm cm}^2\, {\rm s}^{-1}}{K_T}\Bigg)\;
\Bigg(\frac{H_{\rm P}}{10^9 \; {\rm cm}}\Bigg)
\end{equation}
where $K_{\rm T}=100$ cm$^2$s$^{-1}$ is a typical value of the 
conductivity in the white dwarf. Then $P \la 1$ and the gradient 
of carbon mass fraction
\begin{equation}
\Bigg|\frac{dX_{\rm C}}{dr}\Bigg| \simeq 4\times 10^{-12}\;
\Bigg(\frac{v_{\rm crys}}{0.1\,{\rm cm/yr}}\Bigg)\;
\Bigg(\frac{\Delta X_{\rm C}}{0.1}\Bigg)\;
\Bigg(\frac{100 \; {\rm cm}^2 \, {\rm s}^{-1}}{K_T}\Bigg)\;
\Bigg(\frac{1}{{\rm P}}\Bigg)
\end{equation}
is so small that $X_{\rm C}$ varies by less than 1\% in the convective 
region.

We now check that the small superadiabaticity hypothesis is correct, 
i.e.:
\begin{equation}
\chi - k\epsilon \ll \chi
\end{equation}
We first compute $\chi$ with the simplified equation of state described
in \S2. For a crystallizing C/O white dwarf we get

\begin{equation}
\chi \sim  10^{-4}\; 
\Bigg(\frac{H_{\rm P}}{10^9 \, {\rm cm}}\Bigg)\;
\Bigg(\frac{3\times 10^8\, {\rm cm}\;{\rm s}^{-1}}{c_{\rm s}}\Bigg)^{-2}
\Bigg(\frac{\Big|\frac{dX_{\rm C}}{dr}\Big|} {4\times 10^{-12}}\Bigg)
\end{equation}
An upper limit of the superadiabacity can be obtained by assuming that, 
due to the interaction of convection with rotation, the Rossby number
\begin{equation}
R_0 = \frac{v_{\rm conv}}{\omega \; l}
\end{equation}
is equal to unity, where $\omega$ is the angular velocity of the white 
dwarf (Stevenson and Salpeter 1976). Then,
\begin{equation}
\chi - k\epsilon = \Bigg(\frac{H_{\rm P} \omega}{c_{\rm s}}\Bigg)^2 =
3\times 10^{-7}
\Bigg(\frac{H_{\rm P}}{10^9 {\rm cm}}\Bigg)^2
\Bigg(\frac{c_{\rm s}}{3\times 10^8\, {\rm cm}\;{\rm s}^{-1}}\Bigg)^{-2}
\Bigg(\frac{\Pi}{10 \;{\rm h}}\Bigg)
\end{equation}
where $\Pi$ is the rotation period. We therefore conclude that, even 
when rotation is considered, convection is indeed an efficient 
mechanism to redistribute the carbon rich fluid out from the 
crystallization front and that the liquid phase can be considered 
well mixed.

\section{Consequences on white dwarf cooling}

Figure 1 displays the evolution of the energy released per unit mass 
crystallized as the solidification proceeds. For the sake of simplicity 
we have assumed in these calculations that the white dwarf is made of 
an homogeneous 50:50 (by mass) mixture of carbon and oxygen.

\placefigure{fdel}

The energy released near the center of the white dwarf is 
$\epsilon_{\rm g} =3.54 \times 10^{13}$ erg/g and the partial 
contribution of electrons and ions are, respectively $\epsilon_{\rm 
e}= -5.00\times 10^{12}$ erg/g and $\epsilon_{\rm i}= 4.04\times 
10^{13}$ erg/g. The electron term is negative and different from 
zero only due to the mass defect of oxygen ($A_0= 16-3.18\times 
10^{-4}$) relative to carbon. If we had considered other species 
with a higher number of neutrons as compared with protons, as is 
the case of $^{22}$Ne or $^{56}$Fe, the situation would have been 
the reverse. The total energy released during this process is $1.95
\times 10^{46}$ erg.

Since, to a good approximation, the luminosity of a white dwarf 
can be considered to be a function of the temperature of its nearly 
isothermal core, $T_{\rm c}$, it is possible to estimate the delay 
introduced by the solidification as:
\begin{equation}
\Delta t = \int ^{M_{\rm WD}}_0 \frac{\epsilon_{\rm g}(T_{\rm c})}
{L(T_{\rm c})} dm
\end{equation}
where $T_{\rm c}(m)$ is the core temperature when the crystallization
front is located at $m$. The extra time to reach a luminosity $\log 
(L/L_\odot) =-4.5$ is 1.81 Gyr. To compute it we have used the same 
relationship between the luminosity and the core temperature as in 
Segretain et al. (1994). Of course, the total delay essentially depends 
on the transparency of the envelope. Any change in one sense or another 
can amplify or damp the influence of solidification in the cooling of 
white dwarfs and for the moment there are not completely reliable 
envelope models at low luminosities (Mazzitelli 1994).

\section{Conclusions}
We have provided a new formulation of the thermodynamics of 
phase separation upon crystallization that proves that chemical 
differentiation results in a net release of heat that is radiated 
away thus delaying the cooling of the star. This extra heating is 
not due to compressional work, but to the local changes of chemical 
abundances. This formulation is very simple and is suitable for 
introduction in evolutionary codes.

We have also shown that the hypothesis of perfect rehomogeneization 
of the liquid is reasonable and can be used to simplify the problem. 
Notice that if this was not the case, this would result in a decrease 
of the efficiency of the redistribution process as an energy 
source but would not 
invalidate the result that if there is some redistribution there is a 
release of heat.

Finally, we want to emphasize that the total delay introduced by this 
extra source of energy depends on the phase diagrams, on the initial 
chemical compositions and on the transparency of the outer envelope. 
Any change  in these factors can produce noticeable changes in the 
final outcome. 

\noindent
{\em Acknowledgements}
This work has been supported by the DGICYT grants PB94--0111, 
PB94--0827--C02--02, by the CIRIT grant GRQ94--8001 and by the 
AIHF237B.

\newpage

\figcaption{
Energy released by the redistribution of elements per unit of 
crystallized mass (continuous line) and the corresponding time 
delay in the cooling (dotted line) as a function of the solid 
mass (lower panel) and the luminosity (upper panel).
\label{fdel}}

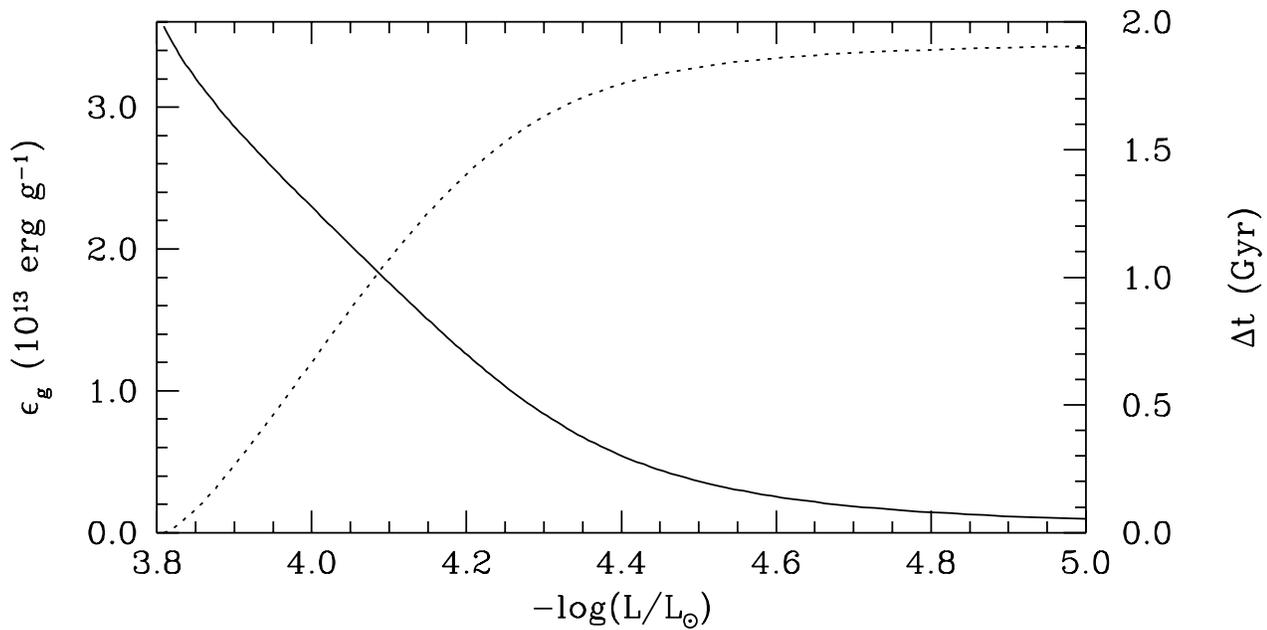
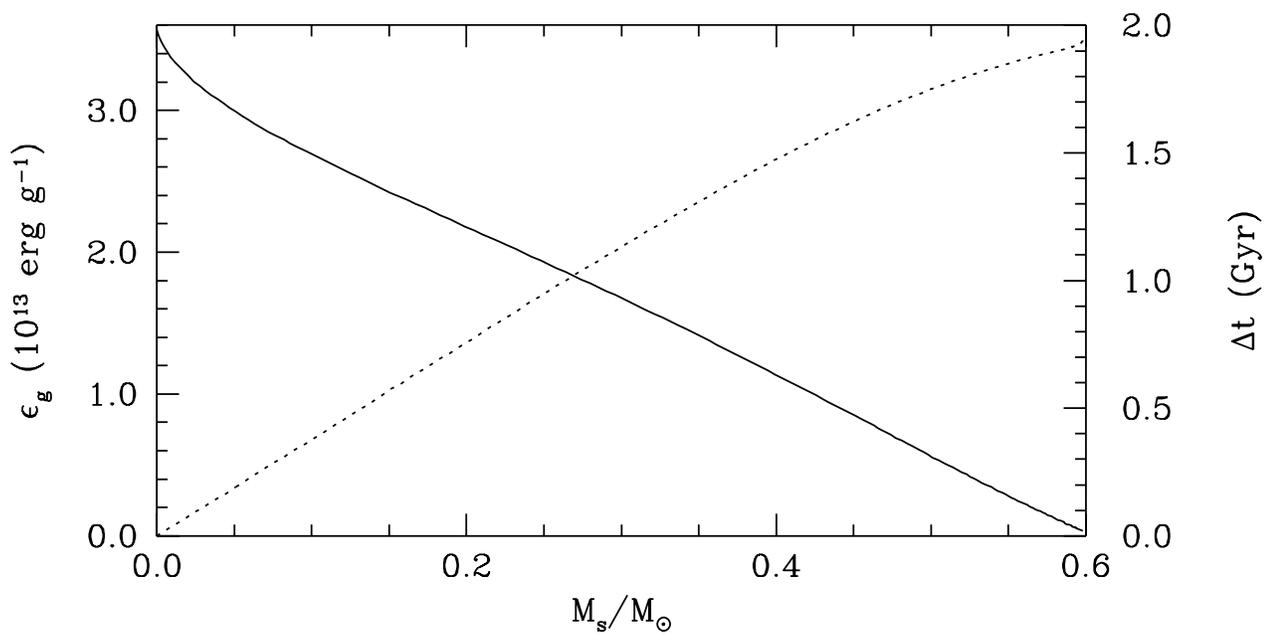
\end{document}